\documentclass[a4paper,10pt]{IEEEtran}
\usepackage[left=1.3cm,right=1.3cm,top=1.82cm, bottom = 4.4cm]{geometry}

\IEEEoverridecommandlockouts
\long\def\comment#1{}

\usepackage{calc} % To reset the counter in the document after title page    

\usepackage{parskip}
\usepackage{algorithm}% http://ctan.org/pkg/algorithm
\usepackage{algpseudocode}% http://ctan.org/pkg/algorithmicx
\usepackage{cite}
\usepackage{amsmath,amssymb,amsfonts}
\usepackage{textcomp}
\usepackage{xcolor}
\usepackage{enumitem}

\usepackage{times}
\usepackage{soul}
\usepackage{url}
\usepackage{ulem}

\usepackage{setspace}
\usepackage{cite}
\usepackage{amsmath,amssymb,amsfonts}
\usepackage{graphicx}
\usepackage{textcomp}
\usepackage{xcolor}
\usepackage[detect-all]{siunitx}

\usepackage{xspace}
\usepackage{subcaption}
\usepackage{bm}
\usepackage{array}
\newcolumntype{P}[1]{>{\centering\arraybackslash}p{#1}}
\newcolumntype{M}[1]{>{\centering\arraybackslash}m{#1}}

\usepackage{algorithm,tabularx}
\makeatletter
\newcommand{\multilines}[1]{%
	\begin{tabularx}{\dimexpr\linewidth-\ALG@thistlm}[t]{@{}X@{}}
		#1
	\end{tabularx}
}
\makeatother

\usepackage{mathtools}

\usepackage{amsthm}
\usepackage{multirow}
\usepackage{color}
\usepackage{epstopdf}
\usepackage{endnotes}
\usepackage{framed}
\usepackage{cool} % for partial derivative
	\usepackage[subnum]{cases}
	\usepackage{multicol}
	\usepackage{tablefootnote}
	
	\DeclareMathOperator*{\argmin}{arg\,min}

	\DeclarePairedDelimiterX{\norm}[1]{\lVert}{\rVert}{#1}
	\DeclarePairedDelimiterX{\abs}[1]{\lvert}{\rvert}{#1}
	\DeclarePairedDelimiterX{\innProd}[1]{\langle}{\rangle}{#1}

	\newcommand{\SumNoLim}[2]{\ensuremath{\sum\nolimits_{#1}^{#2}}}

	\theoremstyle{plain}

	\theoremstyle{definition}

\newcommand{\bU}{\mathbf{U}}
\newcommand{\bX}{\mathbf{X}}

\newcommand{\bV}{\mathbf{V}}

\begin{document}
\newcommand{\pp}[1]{\textcolor{red}{#1}}
\newcommand{\phuc}[1]{\textcolor{black}{#1}}
\newcommand{\forest}[1]{\textcolor{orange}{#1}}

\title{Microsecond Federated SVD on Grassmann Manifold for Real-time IoT Intrusion Detection}

\author{Tung-Anh Nguyen, Van-Phuc Bui, Shashi Raj Pandey, Kim Hue Ta, Nguyen H. Tran, Petar Popovski
\thanks{T.-A Nguyen, N. H. Tran (emails: tung6100@uni.sydney.edu.au,  nguyen.tran@sydney.edu.au) are with School of Computer Science, The University of Sydney, Australia. V.-P Bui, S.R. Pandey, and P. Popovski (emails: \{vpb, srp, petarp\}@es.aau.dk) are with the Department of Electronic Systems, Aalborg University, Denmark. Kim Hue Ta is with School of Electrical and Electronic Engineering, Hanoi University of Science and Technology, Vietnam.}}
% \thanks{This work was supported by the Villum Investigator Grant ``WATER" from the Velux Foundation, Denmark}}

\maketitle
\thispagestyle{empty}
\begin{abstract}
This paper introduces FedSVD, a novel unsupervised federated learning framework for real-time anomaly detection in IoT networks. By leveraging Singular Value Decomposition (SVD) and optimization on the Grassmann manifolds, FedSVD enables accurate detection of both known and unknown intrusions without relying on labeled data or centralized data sharing. Tailored for deployment on low-power devices like the NVIDIA Jetson AGX Orin, the proposed method significantly reduces communication overhead and computational cost. Experimental results show that FedSVD achieves performance comparable to deep learning baselines while reducing inference latency by over $10\times$, making it suitable for latency-sensitive IoT applications. %These results highlight the potential of FedSVD as a lightweight, privacy-preserving, and scalable solution for next-generation intrusion detection systems.
\end{abstract}
% \begin{IEEEkeywords}
%       Federated Learning, Real-time Anomaly Detection, IoT, Singular Value Decomposition, Grassmann Manifold
% \end{IEEEkeywords}

%%%%%%%%%%%%%%%%%%%%%%%%%%%%%%%%%%%%%%%%%%%%%%%%
\vspace*{-5pt}
\section{Introduction}\label{sec:intro}
%%%%%%%%%%%%%%%%%%%%%%%%%%%%%%%%%%%%%%%%%%%%%%%%
The deployment of 5G and beyond wireless networks has enabled transformative applications such as Virtual Reality (VR), eXtended Reality (XR), and large-scale Internet of Things (IoT) systems~\cite{wang2022survey}. These technologies rely on extensive networks of distributed, low-power edge devices that continuously generate and exchange large volumes of data. However, the rapid proliferation of these devices poses significant challenges to system integrity, as anomalies arising from malfunctions, cyber threats, or unexpected data patterns can severely degrade network performance and compromise security. In such a case, the efficacy of traditional centralized anomaly detection methods, such as ~\cite{Al-Fuqaha2015}, which model normal network behaviour and identify deviations indicative of threats, is limited.

Machine learning-based intrusion detection systems (ML-IDS)  offer a promising solution for network defense~\cite{chou2021survey}, typically employing supervised learning approaches from classical classifiers to deep neural networks. However, centralized ML-IDS architectures are increasingly impractical in IoT environments due to data privacy concerns and resource limitations, such as bandwidth and latency constraints. Federated Learning (FL) offers a decentralized alternative by enabling local training while sharing only model parameters~\cite{Belenguer2022}. Despite its advantages, most FL-IDS rely on labeled data, limiting their ability to detect unknown attacks~\cite{Casas2012,Belenguer2022}. Supervised learning models typically incur substantial computational and memory overhead, posing challenges for deployment on resource-constrained IoT devices. In practice, IoT nodes often host multiple ML models to support diverse functionalities, which further exacerbates memory and processing limitations~\cite{zhang2020edge}. This contention for limited resources hinders real-time performance and diminishes the responsiveness of IDS. Additionally, the recent progress in ML has been predominantly driven by deep learning techniques, which require high-performance hardware accelerators such as Graphics Processing Units (GPUs) or Neural Processing Units (NPUs). These requirements significantly increase deployment costs and scalability challenges in large-scale IoT environments. As a result, legacy IoT devices, lacking such specialized hardware, are unable to accommodate modern ML models, leaving them more vulnerable to security threats. These limitations underscore the urgent need for efficient, lightweight, and privacy-preserving anomaly detection approaches, thereby motivating the investigation of unsupervised and low-complexity alternatives.

\begin{figure}[t]
	\centering    
	\includegraphics[width=0.8\columnwidth]{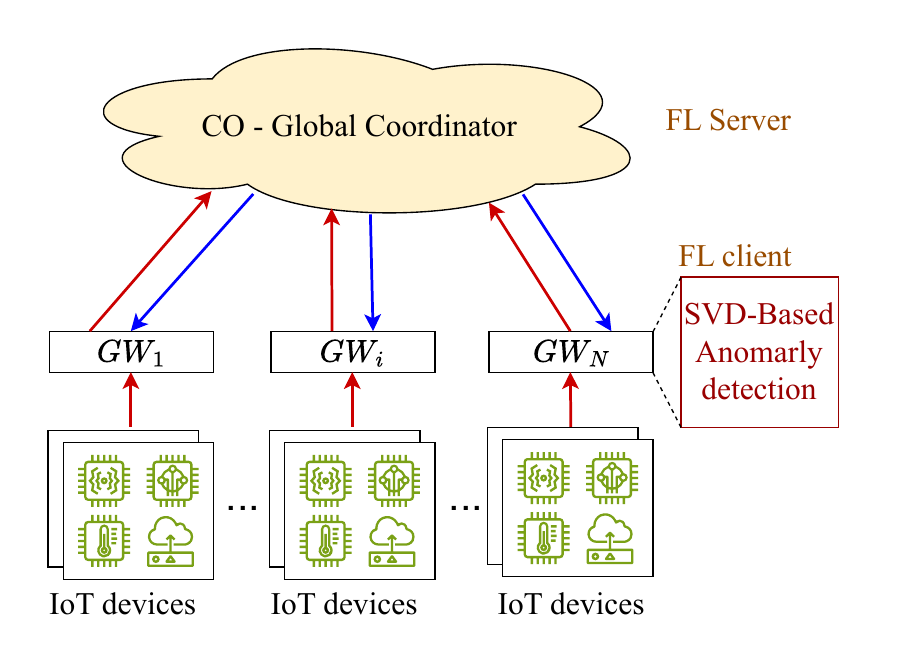}
    % \vspace{-10pt}
	\caption{The visualization of the Federated SVD-based anomaly detection system model.}
	\label{fig:system}
    \vspace{-20pt}
\end{figure}

To tackle these challenges, we introduce \textbf{FedSVD}, a novel unsupervised federated anomaly detection framework leveraging Singular Value Decomposition (SVD) and Grassmann manifold optimization tailored specifically for decentralized IoT networks. \phuc{Compared to neural network-based federated methods, FedSVD significantly reduces communication and storage overhead by exchanging only compact subspace representations. Fig. 1 illustrates this with a toy example, highlighting FedSVD’s lightweight nature and its effectiveness in identifying anomalies without heavy computation. This federated approach, which operates on unlabeled data, can identify both known and previously unseen threats, offering a broader range of generalization than traditional supervised methods.} 
Besides, FedSVD effectively captures the intrinsic geometric structure of high-dimensional network traffic data, providing rapid convergence and accurate anomaly detection without labeled data or centralized data collection~\cite{Patwari2005}. To validate our approach, extensive experiments are performed on the NSL-KDD dataset~\cite{Tavallaee2009}, demonstrating that FedSVD significantly outperforms baseline methods in terms of computational efficiency, and scalability, while keeping high detection accuracy, confirming its practical suitability for resource-limited hardware platforms. 

The main contributions of this work are as follows: $(1)$ We propose FedSVD, a novel FL-based anomaly detection framework leveraging SVD and Grassmann manifold optimization~\cite{Jiayao2018}, specifically designed for decentralized and resource-constrained IoT networks. $(2)$ We develop a distributed optimization algorithm operating on Grassmann manifolds, ensuring efficient and accurate anomaly detection by effectively modeling high-dimensional data. This approach facilitates proactive threat mitigation by efficiently modeling normal behavior and detecting deviations without the need for labeled data or centralized data sharing. $(3)$  Experiments on the NSL-KDD dataset show that FedSVD achieves performance comparable to advanced non-linear baselines while reducing inference latency by over $10\times$, confirm its practical deployment feasibility through implementation on the NVIDIA Jetson AGX Orin platform.

% \pp{PP: In fact, it is not clear from the way it is presented which specific challenge for IoT is solved by FedSVD. Is it lightweight in terms of computation or data storage? This needs to be specific. How is this illustrated on Fig. 1? What can be done with FedSVD w.r.t. what is depicted for this figure? The figure should, ideally, be related to a toy example that clearly illustrates the advantages.} 

% \vspace{-5pt}
%%%%%%%%%%%%%%%%%%%%%%%%%%%%%%%%%%%%%%%%%%%%%%%%
\section{System Model}
%%%%%%%%%%%%%%%%%%%%%%%%%%%%%%%%%%%%%%%%%%%%%%%%
% \subsection{SVD for Centralized dataset}\label{}
% \vspace{-5pt}
This section presents a federated anomaly detection system tailored for interconnected environments such as large-scale IoT and edge computing networks. In these settings, local hubs (e.g., sensors, controllers, or VR/AR headsets) act as FL clients that perform on-device training using their own traffic data, while collaborating with a global server ($CO$) hosted in the cloud.  Illustrated in Fig.~\ref{fig:system}, the communication process between local hubs ($GW$) and the $CO$ operates as: during training, both $CO$ and $GW$ engage in a collaborative process, where low-dimensional representations of the data are computed using SVD. In each training round, $CO$ sends the global model to the $GW$s, which update the model based on their locally collected data. These updates are then aggregated by $CO$ to further refine the global model, repeating until convergence. After training is complete, each $GW$ uses the final global model to autonomously detect anomalies in its local data, without requiring labeled datasets. 
% This federated approach, which operates on unlabeled data, can identify both known and previously unseen threats, offering a broader range of generalization than traditional supervised methods. \pp{This should be explained in the introduction with a clear outline of the advantages of FedSVD.}
Assuming that each $GW$ collects data and sends the those data to $CO$, we use $\mathbf{X} \in \mathbb{R}^{d \times D}$ to denote the fully observed dataset, where $d$ is the data dimension and $D$ is the total number of data points. The dataset $\mathbf{X} = [\mathbf{X}_{1}, \mathbf{X}_{2}, \ldots, \mathbf{X}_{N}]$ is distributed across $N$ clients. Each client $i \in \{1, \ldots, N\}$ contains a local dataset of $D_i$ records, denoted by $\mathbf{X}_i = [\mathbf{x}_{1}, \mathbf{x}_{2}, \ldots, \mathbf{x}_{D_i}] \in \mathbb{R}^{d \times D_i}$, where each $\mathbf{x}_i \in \mathbb{R}^{d}$ represents a single record, and $\sum_{i=1}^{N} D_i = D$. The SVD problem seeks the best rank-$k$ ($k < d$) approximation of the matrix $\mathbf{X}$ by another matrix $\tilde{\mathbf{X}}$ such that the reconstruction error, measured by $\|\mathbf{X} - \tilde{\mathbf{X}}\|_F$, is minimized. Here, $\|\cdot\|_F$ denotes the Frobenius norm. The solution $\tilde{\mathbf{X}}$ is typically based on SVD of $\mathbf{X}$, given by
\begin{IEEEeqnarray}{rll}
\mathbf{X} = \mathbf{U} \mathbf{\Sigma} \bV^\top,
\end{IEEEeqnarray}
where $\mathbf{U} \in \mathbb{R}^{d \times d}$ contains the left singular vectors of $\mathbf{X}$ as its columns, $\mathbf{\Sigma} = \text{diag}(\sigma_1, \ldots, \sigma_d)$ contains the singular values ($\sigma_1 \geq \sigma_2 \geq \ldots \geq \sigma_d \geq 0$), and $\mathbf{V} \in \mathbb{R}^{D \times D}$ contains the right singular vectors of $\mathbf{X}$ as its columns. Note that the columns of $\mathbf{U}$ are orthonormal. The rewrite the closed-form solution for the local dataset $ \tilde{\bX}_i $ is expressed as
\begin{IEEEeqnarray}{rll}\label{SVD_i}
\tilde{\bX}_i = \mathbf{U}_i \mathbf{\Sigma}_i \mathbf{V}_i^\top,
\end{IEEEeqnarray}
where $ \mathbf{U}_i \in \mathbb{R}^{d \times k},  \mathbf{\Sigma}_i \in \mathbb{R}^{k \times k} ,  \mathbf{V}_i \in \mathbb{R}^{D \times k} $. We stress that computing a rank-$k$ approximation $(\mathbf{U}_i, \mathbf{\Sigma}_i \mathbf{V}_i)$ enables dimensionality reduction of the dataset. Specifically, the quantity $\tilde{\mathbf{X}}_i \in \mathbb{R}^{k \times D_i}$ in \eqref{SVD_i} serves as a lower-dimensional representation of $\mathbf{X}_i$ that minimizes the reconstruction error. In a distributed setting, the goal is to learn a common representation across all clients' datasets. This involves minimizing the reconstruction error over all devices while ensuring that the approximated matrices maintain a rank of $k$ and that the local matrices $\mathbf{U}_i$ and $\mathbf{V}_i$ are consistent with global consensus variables $\mathbf{U}$ and $\mathbf{V}$. This problem can be formulated as
\setcounter{equation}{3}
\begin{IEEEeqnarray}{rll}\label{Prob:svd_federated}
		\min_{\mathbf{U}, \{\mathbf{\Sigma}\}_i, \mathbf{V}} \quad & \sum_{i=1}^{N} \|\mathbf{X}_i - \mathbf{U}_i \mathbf{\Sigma}_i (\mathbf{V}_i)^\top\|_F^2 \IEEEyessubnumber\label{Prob:svd_federateda}\\
		\mathrm{s.t.} \quad & \text{rank}(\mathbf{X}_i) = k, \IEEEyessubnumber\label{Prob:svd_federatedb}\\
		& \mathbf{U}_i = \bU, \quad \mathbf{V}_i = \bV, \quad \forall i = 1, \dots, N,\IEEEyessubnumber\label{Prob:svd_federatedc}
\end{IEEEeqnarray}
where the constraints \eqref{Prob:svd_federatedc} enforce consensus among all local matrices $(\mathbf{U}_i, \mathbf{V}_i)$~\cite{boyd2011distributed}.
% As  discussed, SVD seeks to minimize the reconstruction error $|\mathbf{X} - \tilde{\mathbf{X}}|_F$ to capture the most salient features of the original data, allowing for accurate reconstruction from the reduced feature set. For an observed record $\mathbf{x} \in \mathbb{R}^d$ at device $i$, the reconstruction error is given by $|\mathbf{x} - \mathbf{U} \mathbf{\Sigma}_i \mathbf{V}^\top|_F^2$. Considering that anomalies are rare and tend to deviate from normal patterns \cite{Patel2019}, the core principle of SVD-based anomaly detection is that anomalous records will exhibit higher reconstruction errors. In practice, SVD is utilized to model normal behavior, and consistent with prior research \cite{Brauckhoff2009}, an anomaly is flagged when the reconstruction error exceeds a predetermined threshold.
SVD aims to minimize the reconstruction error $\|\mathbf{X} - \tilde{\mathbf{X}}\|_F$ to capture the most salient features of the original data, enabling accurate reconstruction from a reduced feature space. Since anomalies deviate from normal patterns~\cite{Patel2019}, the fundamental principle of SVD-based anomaly detection is that anomalous records yield higher reconstruction errors. Following standard practice~\cite{Brauckhoff2009}, an anomaly is detected when the reconstruction error exceeds a predefined threshold.
% \vspace{-5pt}
\section{Federated SVD on Grassmann Manifold for Network Anomaly Detection}
% \vspace{-5pt}
We propose a federated SVD framework on the Grassmann manifolds for network anomaly detection. The method collaboratively learns shared transformation matrices $\mathbf{U}$ and $\mathbf{V}$ across distributed datasets, while each node computes its local singular values $\mathbf{\Sigma}_i$.

% In this section, we develop a federated SVD framework leveraging the Grassmann manifold  for network anomaly detection. Our approach aims to collaboratively learn common transformation matrices $\mathbf{U}$ and $\mathbf{V}$ across distributed datasets, while computing local singular value matrices $\mathbf{\Sigma}_i$ based on these common transformations. This allows us to capture shared structures in the data while preserving privacy.
% \vspace{-5pt}
\subsection{Problem Reformulation}
We first describe methodology finding common orthonormal matrices $\mathbf{U} \in \mathbb{R}^{d \times k}$ and $\mathbf{V} \in \mathbb{R}^{D \times k}$, shared among all clients, while allowing each client to compute its own $\mathbf{\Sigma}_i \in \mathbb{R}^{k \times k}$ based on their local data. The goal is to minimize the reconstruction error $\|\mathbf{X}_i - \mathbf{U} \mathbf{\Sigma}_i \mathbf{V}^\top\|_F^2$ for each client. Given that $\mathbf{U}$ and $\mathbf{V}$ are orthonormal, the optimal closed-form solution for $\mathbf{\Sigma}_i$ can be derived as
\begin{IEEEeqnarray}{rll}\label{optimalSigma}
\mathbf{\Sigma}_i^* = \mathbf{U}^\top \mathbf{X}_i \mathbf{V}.
\label{optimal_sigma}
\end{IEEEeqnarray}
Detailed derivation can be found in Appendix~A.
% Next, we derive a closed-form solution for $\mathbf{\Sigma}_i$ based on $\mathbf{U}$ and $\mathbf{V}$ as following lemma.
% \begin{lemma}
% \label{lemma:sigma_solution}
% Given orthonormal matrices $\mathbf{U} \in \mathbb{R}^{d \times k}$ and $\mathbf{V} \in \mathbb{R}^{D \times k}$, the optimal $\mathbf{\Sigma}_i$ for local device $i$ is given by
% \begin{IEEEeqnarray}{rll}
% \mathbf{\Sigma}_i^* = \mathbf{U}^\top \mathbf{X}_i \mathbf{V}.\label{optimal_sigma}
% \end{IEEEeqnarray}
% \end{lemma}
% \begin{proof}
% Please see the Appendix A.
% \end{proof}
Substituting $\mathbf{\Sigma}_i^*$ back into the objective \eqref{Prob:svd_federateda}, we can obtain $\|\mathbf{X}_i - \mathbf{U} \mathbf{\Sigma}_i^* \mathbf{V}^\top\|_F^2 = \|\mathbf{X}_i - \mathbf{U} \mathbf{U}^\top \mathbf{X}_i \mathbf{V} \mathbf{V}^\top\|_F^2$.
% \begin{equation}
% \begin{aligned}
% \|\mathbf{X}_i - \mathbf{U} \mathbf{\Sigma}_i^* \mathbf{V}^\top\|_F^2 &= \|\mathbf{X}_i - \mathbf{U} \mathbf{U}^\top \mathbf{X}_i \mathbf{V} \mathbf{V}^\top\|_F^2.
% \end{aligned}
% \end{equation}
With these observations, we mathematically rewrite the  problem \eqref{Prob:svd_federated} as
\setcounter{equation}{5}
\begin{IEEEeqnarray}{rll}\label{Prob:svd_loss_with_constraints_device}
\min_{\mathbf{U}, \mathbf{V}} \quad&f(\mathbf{U}_i, \mathbf{V}_i) = \sum_{i=1}^{N}\|\mathbf{X}_i - \mathbf{U}_i \mathbf{V}_i^\top \mathbf{X}_i^\top \mathbf{U}_i \mathbf{V}_i^\top\|_F^2. \IEEEyessubnumber\\
\mathrm{s.t.} \quad &\text{rank}(\mathbf{X}_i) = k,\IEEEyessubnumber \\
&\mathbf{U}_i\mathbf{U}_i^\top = \mathbf{I}, \quad \mathbf{V}_i\mathbf{V}_i^\top = \mathbf{I},  \forall i = 1, \dots, N.\IEEEyessubnumber\\ 
&\mathbf{U}_i = \bU, \quad \mathbf{V}_i = \bV,  \forall i = 1, \dots, N.\IEEEyessubnumber
\end{IEEEeqnarray}

In the following, we present the Grassmanian manifolds-based FedSVD method, which provides an efficient solution to the problem \eqref{Prob:svd_loss_with_constraints_device} by leveraging the FL framework.

\subsection{Grassmanian manifold-based FedSVD}
% Grassmanian manifold $\mathcal{G}(n,k)$ refers to a space of subspaces embedded in a higher dimensional vector space~\cite{Jiayao2018}.  It can be represented as the collection of all $n \times k$ orthonormal matrices $A$ whose columns span these subspaces
% \begin{equation}\label{eq:grass-constraint}
% 	\mathcal{G}(n, k) = \{ \mathrm{span}(A): A \in \mathbb{R}^{n\times k},\ A^\top A= I_k \}.
% \end{equation} 
The solution depends only on the subspaces spanned by \(U\) and \(V\), not on the specific orthonormal bases themselves. This implies that any two orthonormal matrices \(U\) and \(UQ\), where \(Q\) belongs to the orthogonal group \(O(r)\), which consists of all \(r \times r\) matrices representing rotations or reflections that preserve length and angles, span the same subspace and therefore correspond to equivalent solutions. To eliminate this ambiguity, the optimization is naturally performed over the Grassmann manifold, defined as the space of all \(k\)-dimensional subspaces in \(\mathbb{R}^n\). Formally, the Grassmann manifold \(\mathcal{G}(n,k)\) is given by
\[
\mathcal{G}(n,k) = \{ \mathrm{span}(A) : A \in \mathbb{R}^{n \times k}, \quad A^\top A = I_k \},
\]
which is the set of all \(n \times k\) orthonormal matrices whose columns span these subspaces~\cite{Jiayao2018}.

To perform gradient-based optimization on the Grassmann manifold, we utilize the concept of a tangent space. At a point $A \in \mathcal{G}(n,k)$, the tangent space consists of all matrices $\nabla$ satisfying $A^\top \nabla = 0$.
Given a function $F: \mathcal{G}(n,k) \rightarrow \mathbb{R}$ with Euclidean gradient $F_A$ at $A$, the Riemannian gradient is obtained by projecting $F_A$ onto the tangent space~\cite{sato2019riemannian} as
\begin{equation}
	\nabla_A F = (I_n - A A^\top) F_A.
\end{equation}
To update $A$ along the manifold, we perform a retraction to map the tangent vector back onto the manifold efficiently, often using QR decomposition~\cite{Jiayao2018}
\begin{equation}
	A_{\text{new}} = R(A - \eta \nabla_A F),
\end{equation}
where $\eta$ is the step size and $R$ is the retraction function on the Grassmann manifold.

We employ manifold optimization techniques to solve the problem in \eqref{Prob:svd_loss_with_constraints_device}. Specifically, we perform gradient descent on the Grassmann manifold, using the geometry of the manifold to ensure that updates remain within the feasible set.
The gradient of the objective function with respect to $\mathbf{U}$ and  $\mathbf{V}$ are respectively given by
\begin{align}
    \nabla_{\mathbf{U}} f = -2 \sum_{i=1}^N \left( \mathbf{X}_i - \mathbf{U} \mathbf{U}^\top \mathbf{X}_i \mathbf{V} \mathbf{V}^\top \right) \mathbf{V} \mathbf{V}^\top \mathbf{X}_i^\top.\\
    \nabla_{\mathbf{V}} f = -2 \sum_{i=1}^N \left( \mathbf{X}_i^\top - \mathbf{V} \mathbf{V}^\top \mathbf{X}_i^\top \mathbf{U} \mathbf{U}^\top \right) \mathbf{U} \mathbf{U}^\top \mathbf{X}_i.
\end{align}
These gradients are projected onto the tangent spaces of the Grassmann manifolds to obtain the Riemannian gradients. The updates are then performed using retraction operations to ensure that the updated $\mathbf{U}$ and $\mathbf{V}$ remain on the manifolds.

For a new data point $\mathbf{x}$, the reconstruction error is then calculated as $\epsilon = \|\mathbf{x} - \mathbf{U} \mathbf{U}^\top \mathbf{x} \mathbf{V}^\top \mathbf{V}\|_2$, which is only depended on $ \mathbf{U}$ and $ \mathbf{V}$.

\begin{algorithm}[t]
    % \algsetup{linenosize=\tiny}
    \small
	\caption{Federated SVD on Grassmann manifold (FedSG)}
    \label{algo2}
	\begin{algorithmic}[1]
% 		\State \tbf{Input:} $U$
		\State Randomly initialize $(\bU^0, \bV^0)$ and $(\bU^0_i, \bV_i^0), \forall i$
		\For{$k = 0, \ldots, T - 1$} \Comment{\textit{Global  rounds}}
	%	\State Broadcast $Z^{k}$ to all clients

		\For {client $i = 1, \ldots, N$ in parallel} 
            % \State  $\mathbf{\Sigma}_i^* = { \mathbf{V}^k}^\top \mathbf{X}_i^\top {\mathbf{U}}^k$
 		\State $\bU^{k+1}_i =\underset{\bU}{ \argmin} f(\mathbf{\Sigma}_i^*, \bU, \bV^{k})$
% 		\tcp*[f]{Local rounds x
		\State $\bV_i^{k+1} =\underset{\bV}{ \argmin} f(\mathbf{\Sigma}_i^*, \bU^{k+1}, \bV)$  \Comment{\textit{Lodcal update}}
            \State $\bU^{k+1}_i = R(\bU^{k+1}_i)$
            \State $\bV^{k+1}_i = R(\bV^{k+1}_i)$
		\EndFor
% 		\State Aggregate all $U_i^{k+1}$ from clients
		\State  $\bU^{k+1} = \frac{1}{N} \SumNoLim{i=1}{N} \bU_i^{k+1}$\Comment{\textit{Global update}}
            \State $\bV^{k+1} = \frac{1}{N} \SumNoLim{i=1}{N} \bV_i^{k+1}$
            \State  $\bU^{k+1} = R(\bU^{k+1})$ $\leftarrow$ projected to the Grassmann manifolds
            \State $ \bV^{k+1} = R(\bV^{k+1})$
		\State Broadcasts $(\bU^{k+1}, \bV^{k+1})$ to all clients %https://www.overleaf.com/project/60c1859569c4be9bf761503d
		% \For {client $i = 1, \ldots, N$ in parallel} \Comment{\textit{Local update}}
		% \State   $Y_i^{k+1} = Y_i^k + \rho \BigP{U_i^{k+1} - Z^{k+1}} $ %\Comment{Local computation}
		% \State   $T_i^{k+1} = T_i^k + \rho \, h_i(U_i^{k+1})$ %\Comment{Local update}
		% \EndFor
		\EndFor
	\end{algorithmic}
\end{algorithm}

\subsection{Complexity and Scalability Analysis}
The computational complexity of FedSVD involves both local and global update procedures. At each local client, the primary cost arises from computing the SVD of the local dataset $\mathbf{X}_i \in \mathbb{R}^{d \times D_i}$ with complexity $\mathcal{O}(\min(d^2 D_i, d D_i^2))$, and updating the matrices $\mathbf{U}_i$ and $\mathbf{V}_i$ on the Grassmann manifold, which incurs an additional cost of $\mathcal{O}(d k^2 + D_i k^2)$. Thus, the total local computational complexity per client per iteration is $\mathcal{O}(\min(d^2 D_i, d D_i^2) + d k^2 + D_i k^2)$. At the server, aggregation requires averaging the matrices $\mathbf{U}_i$ and $\mathbf{V}_i$ from $N$ clients, each of size $d \times k$ and $D \times k$, respectively, resulting in a global update complexity of $\mathcal{O}(N d k + N D k)$.

The communication overhead per client per iteration involves transmitting the updated matrices $\mathbf{U}_i$ and $\mathbf{V}_i$, with $d \times k$ and $D_i \times k$ elements, leading to a communication cost of $\mathcal{O}(k(d + D_i))$. In comparison to FL approaches based on neural networks, where the communication overhead scales with the total number of model parameters $\mathcal{O}(w)$ with $w$ being the number of neural network parameters, FedSVD offers significantly reduced overhead, making it suitable for constrained-bandwidth IoT and edge environments. 
% \pp{PP: Is there a way to mention this in relation to Fig. 1 in the introduction?}

% Finally, FedSVD demonstrates scalability with respect to the network size $N$ and the dimensional parameters $d$ and $k$, leveraging the low-rank assumption $k \ll d$ to maintain low computational and communication complexity even in large-scale distributed IoT systems.
%%%%%%%%%%%%%%%%%%%%%%%%%%%%%%%%%%%%%%%%%%%%%%%%

\section{Experiement Results}
%%%%%%%%%%%%%%%%%%%%%%%%%%%%%%%%%%%%%%%%%%%%%%%%
In this section, we detail the experimental setup employed to evaluate the effectiveness of the proposed methods. We first describe the dataset, parameter configurations, and performance evaluation metrics used throughout the experiments. To demonstrate the advantages of our approach, we conduct a comparative analysis between FedSVD and several baseline methods within the context of an IDS deployment.
\subsection{Experiment settings}
Our implementation is based on PyTorch. To demonstrate the practical feasibility of the methods on edge devices, we conduct experiments on an NVIDIA Jetson AGX Orin Developer Kit. The platform is equipped with a 12-core ARM Cortex-A78AE v8.2 CPU, a 2048-core Ampere GPU with 64 Tensor Cores, and 64 GB LPDDR5 memory. The system runs Ubuntu 20.04.6 LTS with JetPack 5.1.2 (L4T 35.4.1) and operates under a 15W power mode. The software environment includes CUDA 11.4, cuDNN 8.6.0, and TensorRT 8.5.2.
\subsubsection{Datasets and Metrics}
\label{Sec:dataset}

To evaluate the effectiveness of our approach, experiments are conducted using the NSL-KDD dataset~\cite{Tavallaee2009}. NSL-KDD is a benchmark in network intrusion detection system (NIDS) research, consisting of records that capture connection and traffic characteristics, such as connection duration, protocol type, and transmitted bytes. Each record is labeled as either normal traffic or one of four intrusion types: Denial of Service (DoS), Probing (Probe), User to Root (U2R), or Remote to Local (R2L). Following the procedure proposed in~\cite{nguyen2023federated}, the dataset is divided into a training set with 125,973 records and a testing set with 22,544 records. The testing set includes 17 novel attack types not present in the training set , facilitating the evaluation of FedSVD’s capability to detect never-before-seen attacks.
% The testing set includes 17 novel attack types not present in the training set as depicted in Fig.~\ref{fig:attack_types}, facilitating the evaluation of FedSVD’s capability to detect never-before-seen attacks.

Within the training set, normal traffic accounts for 67,343 samples (53.46\%), while DoS, Probe, R2L, and U2R attacks contribute 45,927 (36.46\%), 11,656 (9.25\%), 995 (0.79\%), and 52 (0.04\%) records, respectively. In the test set, normal records comprise 9,711 instances (43.08\%), whereas DoS, Probe, R2L, and U2R attacks are represented by 7,458 (33.08\%), 2,421 (10.74\%), 2,754 (12.22\%), and 200 (0.89\%) records, respectively.
% \begin{figure}[t]
% 	\centering    
% 	\includegraphics[width=\columnwidth]{figs/data.png}
%     \vspace{-15pt}
% 	\caption{Types of Attacks.}
% 	\label{fig:attack_types}
%     \vspace{-20pt}
% \end{figure}
In line with standard practice ~\cite{nguyen2023federated}, we use 34 features for model training. Reflecting the observation that newly deployed IoT devices generally exhibit only legitimate communication during initial operation~\cite{Nguyen2019}, we exclusively utilize the 67,343 normal samples from the training set to construct a benign traffic profile. Model evaluation is carried out using five performance metrics: Accuracy (Acc), Precision (Pre), True Positive Rate (TPR or Recall), False Positive Rate (FPR), and F1-score (F1). In the context of anomaly detection, TPR quantifies the proportion of correctly identified malicious samples, while FPR measures the rate at which benign samples are misclassified as malicious. An effective anomaly detection system strives to maximize TPR while maintaining a low FPR to ensure detection robustness and reliability. Among the selected metrics, F1-score serves as a comprehensive indicator by blending Precision and TPR into a unified measure~\cite{Belenguer2022}.

For unsupervised anomaly detection, a reconstruction error is computed for each sample based on the learned normal profile. Records exhibiting reconstruction errors exceeding a predefined threshold are flagged as anomalies. Specifically, the detection threshold is selected as the $\rho$-th percentile of the sorted reconstruction error distribution. To comprehensively characterize the trade-off between TPR and FPR under varying threshold settings, we utilize the Receiver Operating Characteristics (ROC) curve~\cite{Brauckhoff2009}.

\subsubsection{Baselines}
\begin{figure*}[h]
	\centering
	\begin{subfigure}{.33\textwidth}
		\centering
		\includegraphics[width=\textwidth]{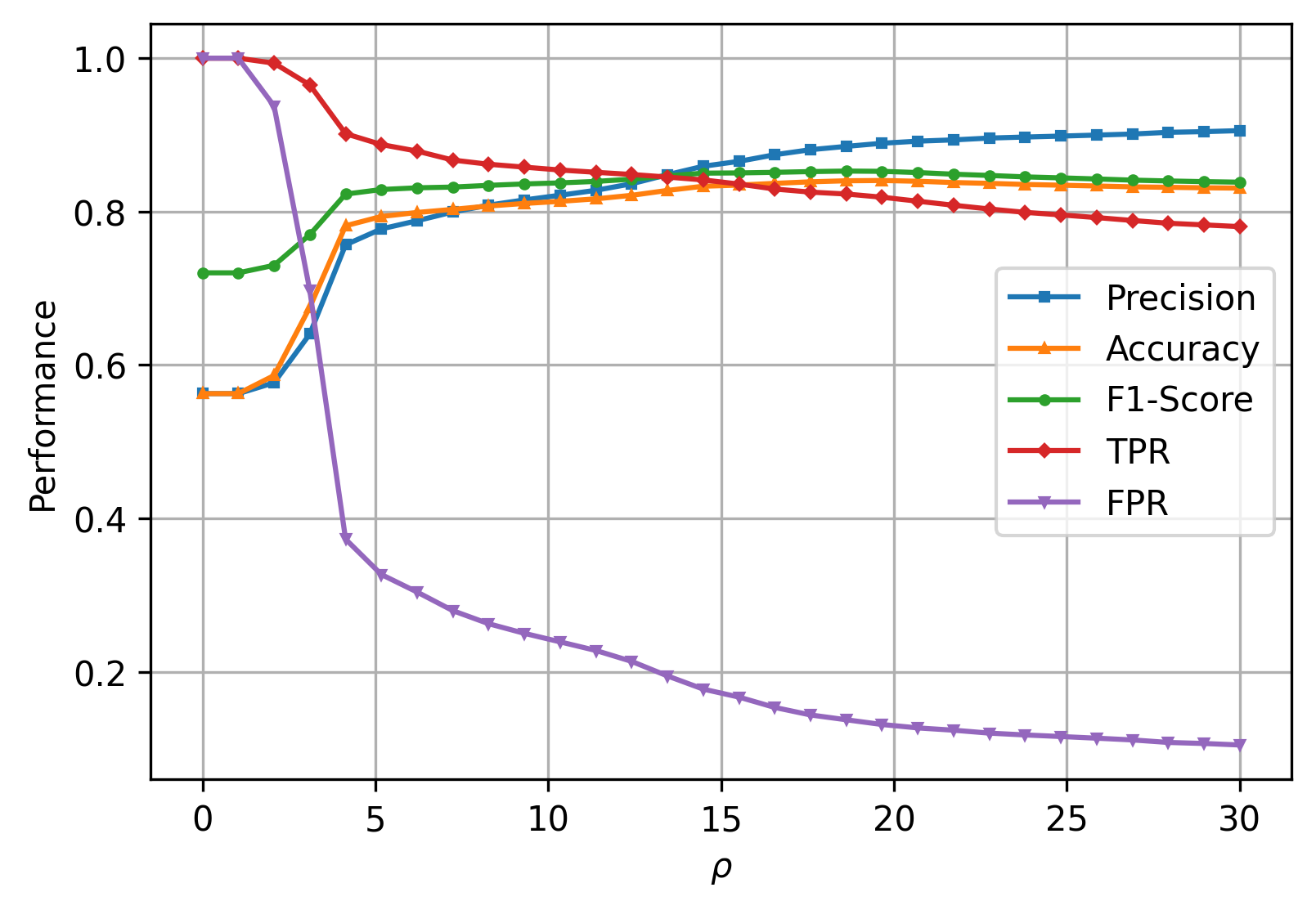}
        %\vspace{-15pt}
            \caption{}
		\label{fig:roc1}
	\end{subfigure}%
 	\begin{subfigure}{.33\textwidth}
		\centering
		\includegraphics[width=\textwidth]{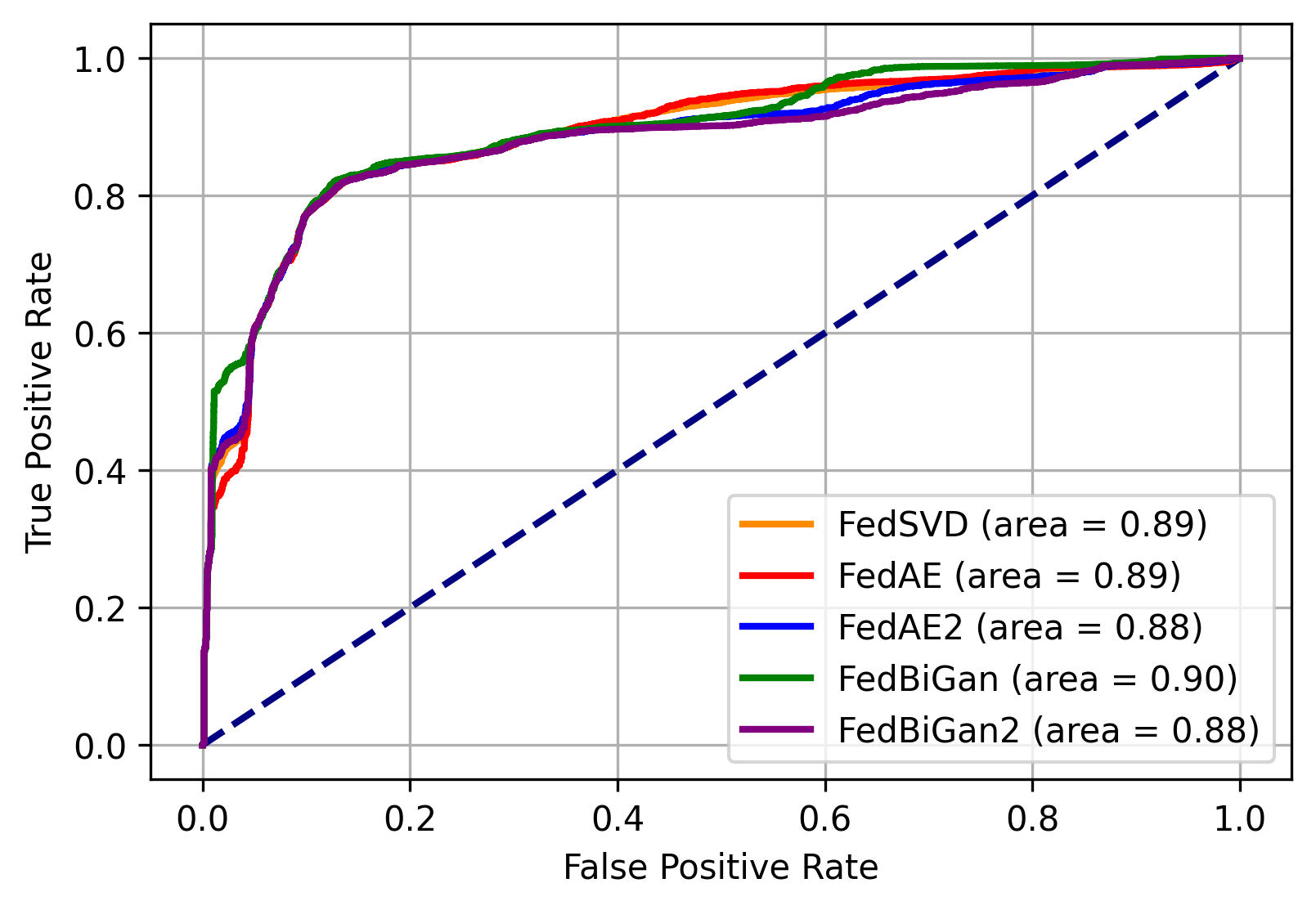}
        %\vspace{-15pt}
            \caption{}
		\label{fig:roc2}
	\end{subfigure}
 	\begin{subfigure}{.33\textwidth}
		\centering
		\includegraphics[width=\textwidth]{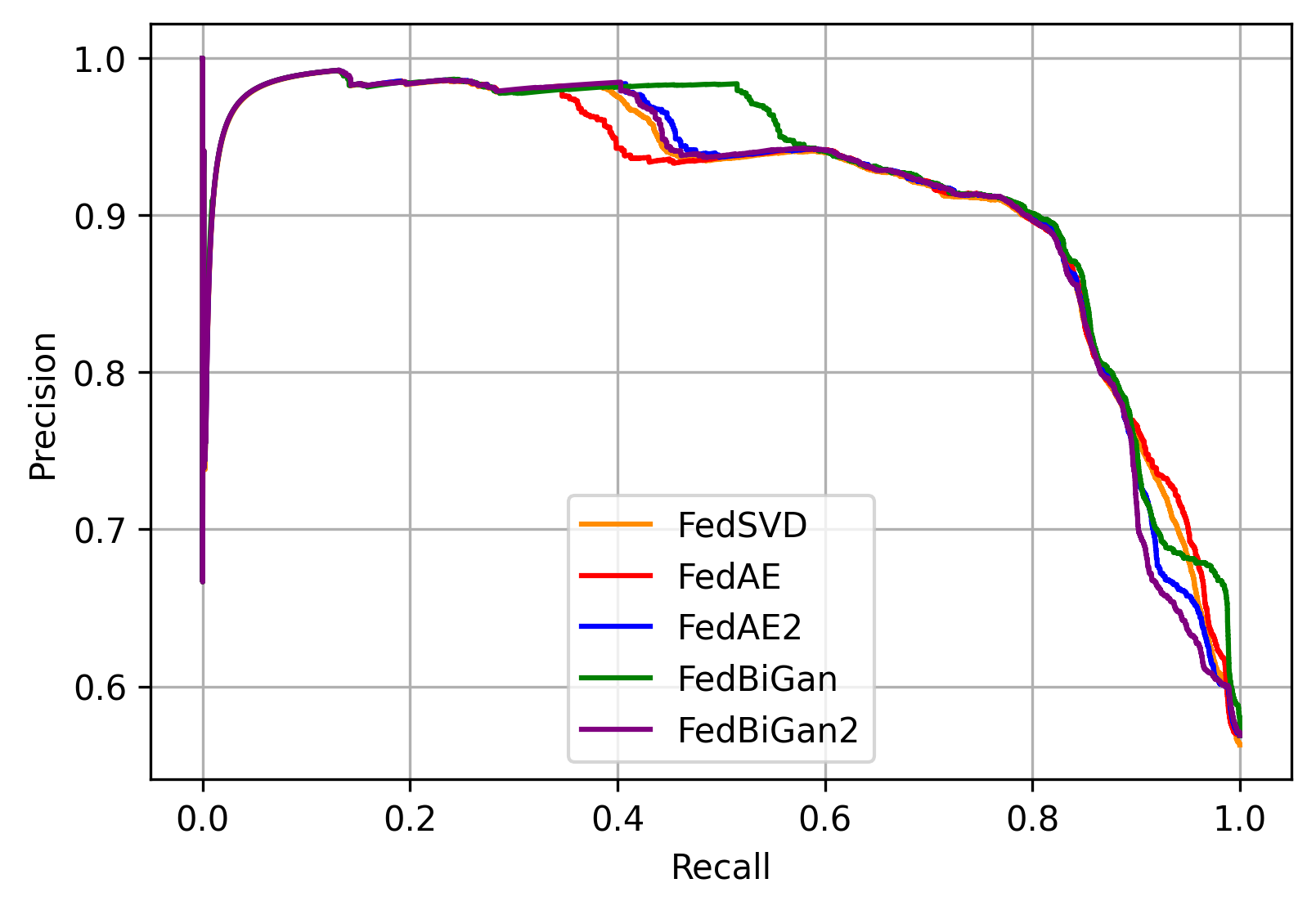}
        %\vspace{-15pt}
            \caption{}
		\label{fig:pre1}
	\end{subfigure}%
 % 	\begin{subfigure}{.24\textwidth}
	% 	\centering
	% 	\includegraphics[width=\textwidth]{figures/ton_prerecall_revise.pdf}
 %        %\vspace{-15pt}
 %            \caption{}
	% 	\label{fig:pre2}
	% \end{subfigure}
	\caption{(a) Performance of FedSVD with different threshold $\rho$, (b) ROC curve and (c) Precision-Recall curve}
  	\label{fig:roc_pre}
    \vspace{-15pt}
\end{figure*}

\begin{figure*}[h]
	\centering
	\begin{subfigure}{.33\textwidth}
		\centering
		\includegraphics[width=\textwidth]{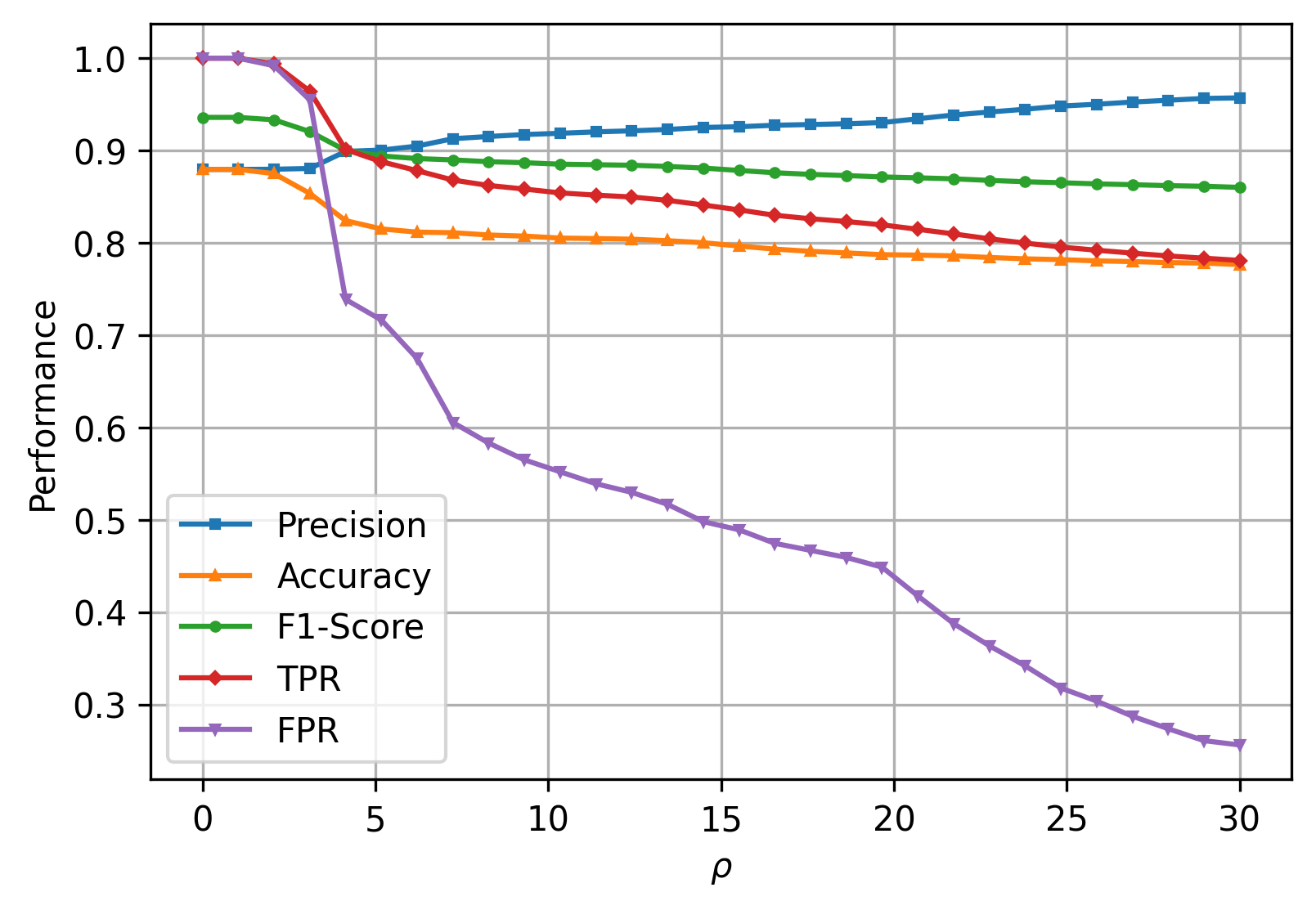}
        %\vspace{-15pt}
            \caption{}
		\label{fig:roc1}
	\end{subfigure}%
 	\begin{subfigure}{.33\textwidth}
		\centering
		\includegraphics[width=\textwidth]{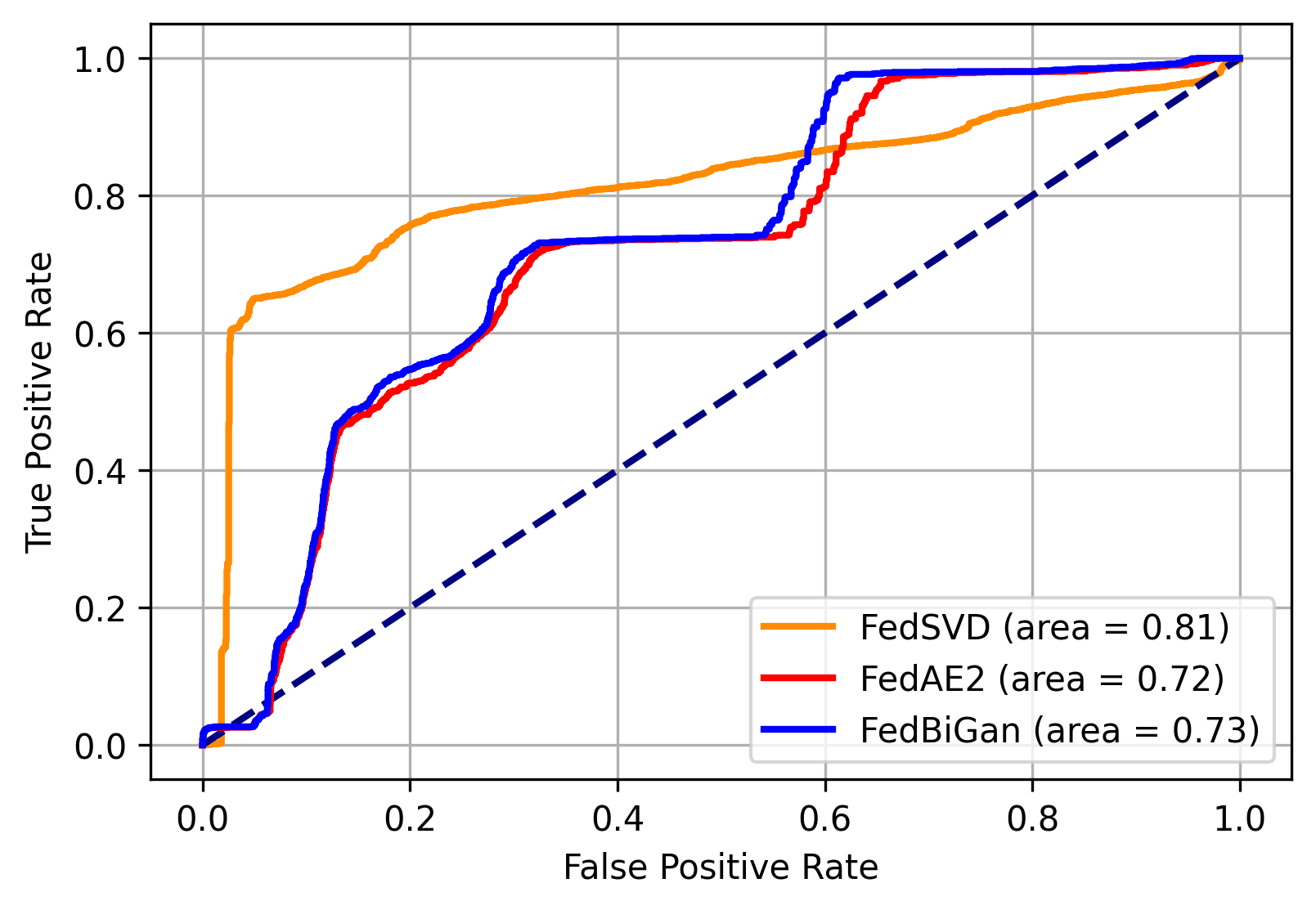}
        %\vspace{-15pt}
            \caption{}
		\label{fig:roc2}
	\end{subfigure}
 	\begin{subfigure}{.33\textwidth}
		\centering
		\includegraphics[width=\textwidth]{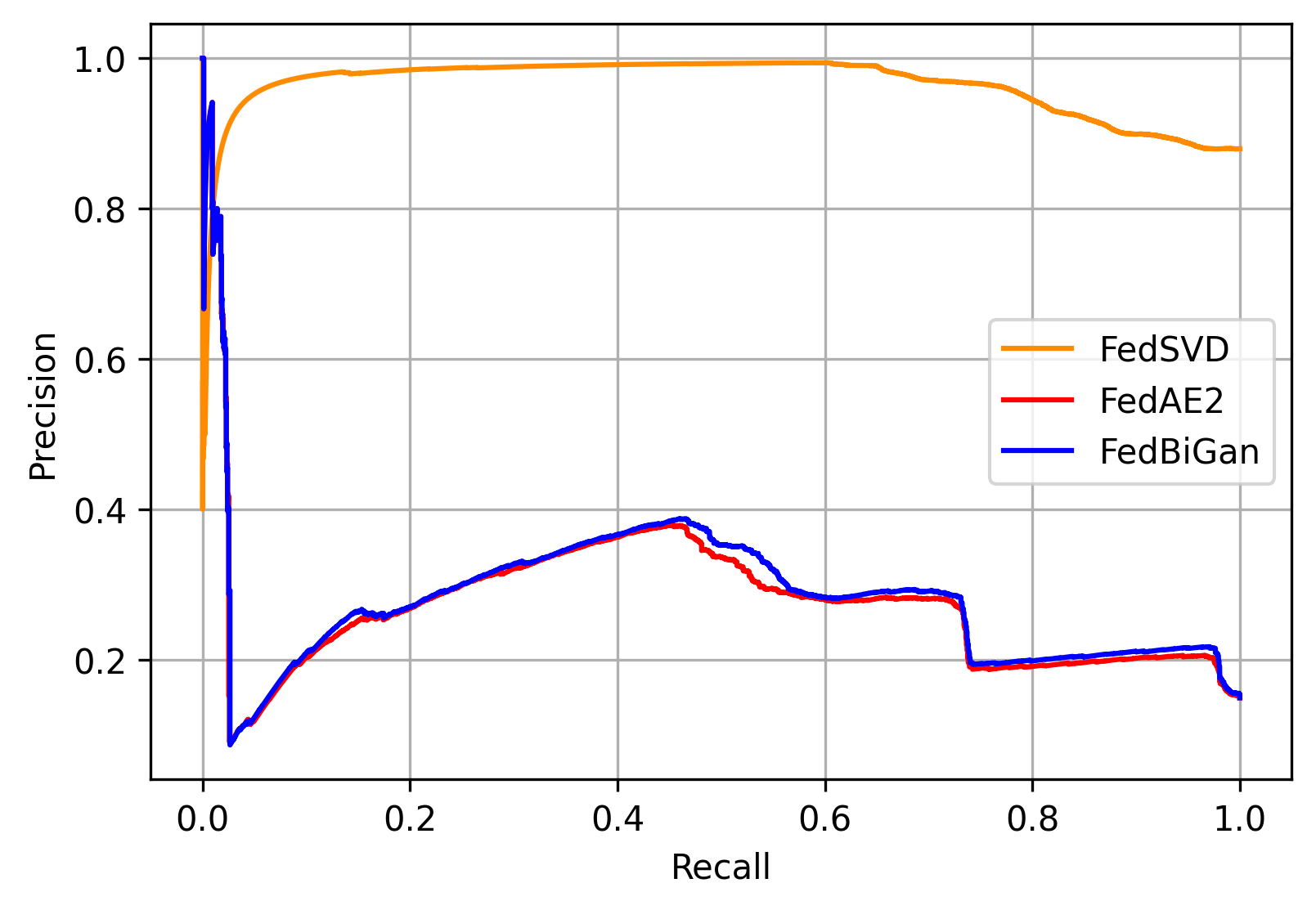}
        %\vspace{-15pt}
            \caption{}
		\label{fig:pre1}
	\end{subfigure}%
 % 	\begin{subfigure}{.24\textwidth}
	% 	\centering
	% 	\includegraphics[width=\textwidth]{figures/ton_prerecall_revise.pdf}
 %        %\vspace{-15pt}
 %            \caption{}
	% 	\label{fig:pre2}
	% \end{subfigure}
	\caption{(a) Performance of FedSVD with different threshold $\rho$ over R2L and U2R attacks, (b) ROC curves over R2L and U2R attacks and (c) Precision-Recall curves over R2L and U2R attacks}
  	\label{fig:unseen_data_perf}
    \vspace{-15pt}
\end{figure*}
% The baseline methods used for comparison are summarized as follows:
% \begin{enumerate}[label=\alph*)]
%     \item \textit{Self-learning SVD}: a decentralized method where each client independently trains a SVD model on its local data. Anomalies are detected by reconstructing samples via low-rank approximations and computing reconstruction errors, with deviations above a threshold indicating abnormality.
%     \item \textit{AutoEncoder (AE)}: an unsupervised encoder-decoder neural network that compresses input data into latent representations and reconstructs them. Anomalies are identified by large reconstruction discrepancies, reflecting deviation from normal data patterns~\cite{ae1}.
%     \item \textit{Bidirectional Generative Adversarial Network (BiGAN)}: a generative model comprising a generator, a discriminator, and an encoder. During training, the encoder maps data to latent representations, and the generator reconstructs data from these codes. The discriminator distinguishes real samples from generated ones, enabling anomaly detection through high reconstruction errors when trained only on normal data~\cite{bigan1}.
% \end{enumerate}
The unsupervised baseline methods used for comparison are summarized as: \textit{(1) Self-learning SVD}: a decentralized method where each client independently trains a SVD model on its local data. Anomalies are detected by reconstructing samples via low-rank approximations and computing reconstruction errors, with deviations above a threshold indicating abnormality.  \textit{(2) AutoEncoder (AE)}: an unsupervised encoder-decoder neural network that compresses input data into latent representations and reconstructs them. Anomalies are identified by large reconstruction discrepancies, reflecting deviation from normal data patterns~\cite{ae1}.  \textit{(3)  Bidirectional Generative Adversarial Network (BiGAN)}: a generative model comprising a generator, a discriminator, and an encoder. During training, the encoder maps data to latent representations, and the generator reconstructs data from these codes. The discriminator distinguishes real samples from generated ones, enabling anomaly detection through high reconstruction errors when trained only on normal data~\cite{bigan1}. 
All baseline models are adapted to the federated setting using the FedAvg~\cite{mcmahan2017communication}. We construct FedAE and FedBiGAN using two-layer encoder-decoder structures, and FedAE-2 and FedBiGAN-2 using four-layer architectures to capture more complex data distributions. For self-learning SVD, each client independently performs anomaly detection, and global performance is assessed by aggregating local results.

\subsubsection{Federated Setting}

In the IDS deployment, local IoT devices collect traffic data as access clients to train anomaly detection models. To simulate non-i.i.d. distributions, the training set is partitioned into 100 subsets based on the 'dst\_bytes' feature. FedSVD trains models locally without sharing raw data, and evaluation is performed on the full test set, which includes unknown attacks. Features are normalized using z-score scaling computed from local training data.

We set $k=3$ selected via grid search for optimal performance and $N=100$ clients. At each communication round, $|S_t| = 20\%$ of clients are randomly sampled. Global and local training rounds are fixed at $T=200$ and $C=5$, respectively. All tasks share consistent hyper-parameter settings.

\begin{table}[t]
	\centering
	\caption{Detection Performance (\%) over NSL-KDD.}
	\label{tab:performance}
	\begin{tabular}{|l|c|c|c|c|c|}
		\hline
		\textbf{Method} & \textbf{Acc} & \textbf{Pre} & \textbf{TPR} & \textbf{FPR} & \textbf{F1} \\ 
		\hline
		\hline
		Self-SVD   & 52.91  & 55.23  & 91.01  & 11.88 &  68.73  \\ 
		\hline
		FedSVD (ours)    & \textbf{82.12} & \textbf{88.95} & \textbf{82.29} & \textbf{13.75}     & \textbf{85.28}  \\ 
		\hline
		FedAE      & 82.91   & 90.08  & 77.72  & 10.41 & 83.79  \\
		\hline
		FedAE-2      & 83.69   & 87.82  & 82.81  & 15.18 & 85.25  \\
		\hline
		FedBiGAN     & 83.69   & 86.76  & 84.44  & 16.92 & 85.62  \\
		\hline
		FedBiGAN-2     & 83.27   & 89.98  & 79.43  & 11.66 & 84.38  \\
		\hline
	\end{tabular}
\end{table}
% \vspace{-15pt}
\subsection{Experimental Results}
\subsubsection{Detection Performance}
\label{sec:detection_performance}
As defined in Sec.~\ref{Sec:dataset}, the threshold $\rho$ governs the trade-off between FPR and TPR in anomaly detection. The performance of FedSVD across different thresholds is shown in Fig.~\ref{fig:roc_pre}(a). Lower thresholds yield higher TPR at the expense of increased FPR, while higher thresholds reduce FPR but lower the detection rate. As observed, the F1-score of FedSVD exceeds 80\% once $p$ surpasses 10, and the FPR falls below 20\%, indicating stable and effective detection. For practical deployments prioritizing a low FPR, a larger threshold is preferable. Accordingly, we set $\rho=18$ for all subsequent experiments and baseline comparisons reported in Table~\ref{tab:performance}.

We observe that FedSVD achieves an F1-score of 85.28\%, which is comparable to the best-performing deep model, FedBiGAN, with 85.62\%. Although FedAE-2 and FedBiGAN attain slightly higher F1-scores, the performance gap remains narrow (within 0.4\%), highlighting the competitive capability of FedSVD despite its simpler model structure. Compared to the decentralized Self-learning SVD baseline, which only achieves an F1-score of 68.73\% and an accuracy of 52.91\%. In terms of robustness, FedSVD maintains a TPR of 82.29\% with a FPR of 13.75\%, outperforming FedAE-2 and FedBiGAN in balancing detection and false alarms. This stable trade-off is critical in IDS deployments, where minimizing false alarm rates is essential for practical system adoption. 

The detection capability of FedSVD is further assessed through the ROC and Precision-Recall (PR) curves shown in Fig.~\ref{fig:roc_pre}(b) and Fig.~\ref{fig:roc_pre}(c). FedSVD achieves an AUC of 0.89, matching FedAE and slightly below FedBiGAN at 0.90, while outperforming FedAE2 and FedBiGAN2 at 0.88. This indicates that FedSVD maintains a high TPR with controlled FPR thresholds. In the PR curves, FedSVD achieves competitive precision across varying recall levels, particularly in the mid-recall range where detection is more challenging.

Unlike most of the DoS and Probe attacks, the U2R and R2L attacks exhibit stealthy behavior that closely resembles benign traffic, lacking frequent sequential patterns~\cite{Tavallaee2009}. This makes them significantly harder to detect, particularly in the presence of unseen attack types during inference. To evaluate the robustness of FedSVD under such challenging conditions, Fig.~\ref{fig:unseen_data_perf} compares performance of FedSVD with two of the best-performing baselines, FedAE2 and FedBiGAN, on the detection of U2R and R2L attacks. The ROC curve shows that FedSVD achieves an AUC of 0.81, outperforming both FedAE2 (0.72) and FedBiGAN (0.73). This margin indicates a substantial improvement in distinguishing attack traffic from normal behavior in FPR regions. Moreover, as observed in the PR curve, FedSVD maintains a consistently higher precision across nearly all recall levels, suggesting superior resilience to false alarms while maintaining high sensitivity. 
% It is also worth noting that FedSVD achieves this improvement while preserving a fully unsupervised and federated training process, providing a practical trade-off between performance and privacy in intrusion detection deployments.
\subsubsection{Model Efficiency}
\begin{figure}[!t]
	\centering    
	\includegraphics[width=\columnwidth]{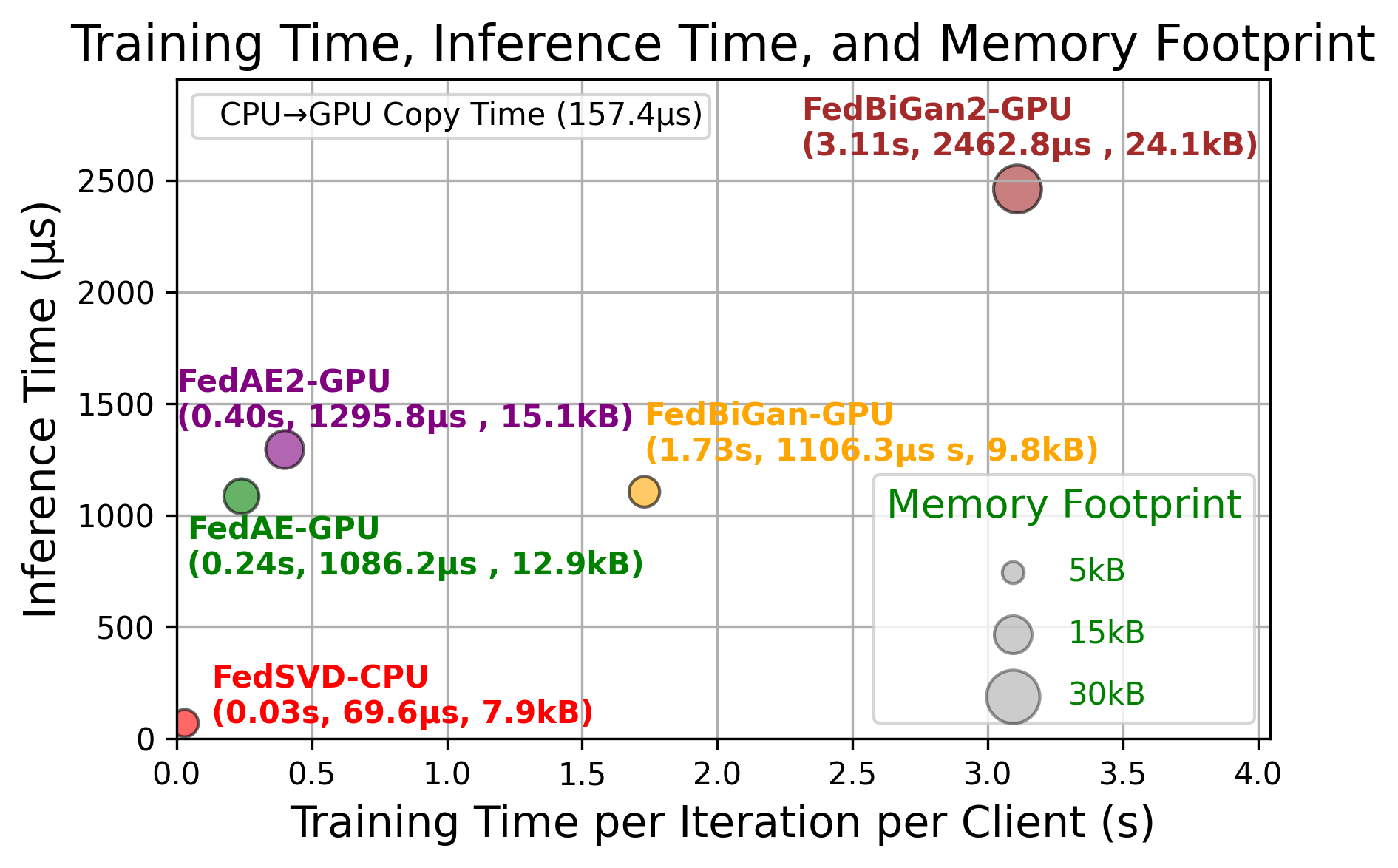}
    % \vspace{-15pt}
	\caption{Model efficiency.}
	\label{fig:model_efficiency}
    % \vspace{-15pt}
\end{figure}

The efficiency of FedSVD the model, corresponding to the detection performance reported in Sec.~\ref{sec:detection_performance}, is illustrated in Fig.~\ref{fig:model_efficiency}. FedSVD achieves the lowest training time of 0.03\,s per iteration at the client side and an inference time of 69.9\,µs per data sample, while operating solely on CPU. In contrast, deep learning-based baselines such as FedAE2, FedBiGAN, and FedBiGAN2 rely on GPU acceleration, yet require between 0.24\,s and 3.11\,s per training round at the client side and over 1100\,µs for inference.  It is also noteworthy that GPU-based models incur additional overhead due to data transfer from CPU to GPU, further increasing end-to-end latency. Moreover, FedSVD exhibits the smallest memory footprint of 7.9\,kB, significantly lower than that of the deep models.

% These results demonstrate that FedSVD not only provides real-time prediction capability with minimal computational overhead but also eliminates the need for specialized hardware. 
%%%%%%%%%%%%%%%%%%%%%%%%%%%%%%%%%%%%%%%%%%%%%%%%
\section{Conclusions}
%%%%%%%%%%%%%%%%%%%%%%%%%%%%%%%%%%%%%%%%%%%%%%%%
% \vspace*{-10pt}
This paper proposed FedSVD, an unsupervised federated anomaly detection framework that combines SVD and Grassmann manifold optimization for real-time IoT intrusion detection. FedSVD enables decentralized learning without requiring labeled data, making it well-suited for resource-constrained IoT devices. Experimental results on the NSL-KDD dataset show that FedSVD achieves competitive detection performance with significantly lower inference latency and memory usage compared to deep learning baselines. Its efficient and lightweight design ensures practical deployment on platforms such as the NVIDIA Jetson AGX Orin.

% \vspace{-15pt}
\section*{Appendix A}\label{appendix:sigma_derivation}
% \vspace{-5pt}
Consider the optimization objective for the $i$-th client, given by $\|\mathbf{X}_i - \mathbf{U} \mathbf{\Sigma}_i \mathbf{V}^\top\|_F^2$.
Expanding the Frobenius norm and expressing the terms via the trace operator yields $\|\mathbf{X}_i - \mathbf{U} \mathbf{\Sigma}_i \mathbf{V}^\top\|_F^2 = \mathrm{Tr}\left((\mathbf{X}_i - \mathbf{U} \mathbf{\Sigma}_i \mathbf{V}^\top)^\top(\mathbf{X}_i - \mathbf{U} \mathbf{\Sigma}_i \mathbf{V}^\top)\right) = \mathrm{Tr}(\mathbf{X}_i^\top \mathbf{X}_i) - 2\,\mathrm{Tr}(\mathbf{X}_i^\top \mathbf{U} \mathbf{\Sigma}_i \mathbf{V}^\top) + \mathrm{Tr}(\mathbf{V} \mathbf{\Sigma}_i^\top \mathbf{U}^\top \mathbf{U} \mathbf{\Sigma}_i \mathbf{V}^\top).$
% \begin{align}
% \|\mathbf{X}_i - \mathbf{U} \mathbf{\Sigma}_i \mathbf{V}^\top\|_F^2 &= \mathrm{Tr}\left((\mathbf{X}_i - \mathbf{U} \mathbf{\Sigma}_i \mathbf{V}^\top)^\top(\mathbf{X}_i - \mathbf{U} \mathbf{\Sigma}_i \mathbf{V}^\top)\right) \nonumber\\
% &= \mathrm{Tr}(\mathbf{X}_i^\top \mathbf{X}_i) - 2\,\mathrm{Tr}(\mathbf{X}_i^\top \mathbf{U} \mathbf{\Sigma}_i \mathbf{V}^\top) \\
% &\quad + \mathrm{Tr}(\mathbf{V} \mathbf{\Sigma}_i^\top \mathbf{U}^\top \mathbf{U} \mathbf{\Sigma}_i \mathbf{V}^\top).\nonumber
% \end{align}
Since $\mathbf{U}$ and $\mathbf{V}$ are orthonormal, we simplify the third term elegantly as $\mathrm{Tr}(\mathbf{V} \mathbf{\Sigma}_i^\top \mathbf{\Sigma}_i \mathbf{V}^\top) = \mathrm{Tr}(\mathbf{\Sigma}_i^\top \mathbf{\Sigma}_i \mathbf{V}^\top \mathbf{V}) = \mathrm{Tr}(\mathbf{\Sigma}_i^\top \mathbf{\Sigma}_i)$. 
% \begin{align}
% \mathrm{Tr}(\mathbf{V} \mathbf{\Sigma}_i^\top \mathbf{\Sigma}_i \mathbf{V}^\top) = \mathrm{Tr}(\mathbf{\Sigma}_i^\top \mathbf{\Sigma}_i \mathbf{V}^\top \mathbf{V}) = \mathrm{Tr}(\mathbf{\Sigma}_i^\top \mathbf{\Sigma}_i).
% \end{align}
Taking the derivative of the objective with respect to $\mathbf{\Sigma}_i$ and employing matrix calculus identities, we have
\begin{align}
\frac{\partial}{\partial \mathbf{\Sigma}_i}\|\mathbf{X}_i - \mathbf{U} \mathbf{\Sigma}_i \mathbf{V}^\top\|_F^2 &= -2\mathbf{U}^\top \mathbf{X}_i \mathbf{V} + 2\mathbf{\Sigma}_i(\mathbf{V}^\top \mathbf{V}).
\end{align}
Given the orthonormal constraint on $\mathbf{V}$, setting this derivative equal to zero yields the optimal $\mathbf{\Sigma}_i^*$ as in \eqref{optimalSigma}.
% \vspace{-10pt}
% \setstretch{0.8}
\bibliographystyle{IEEEtran}
\bibliography{Journal} % Path to your References.bib file
% \vspace{-5pt}
\end{document}